\documentclass[letterpaper,twocolumn]{article}

\usepackage{color}
\usepackage{fullpage}

\definecolor{typicalblue}{rgb}{0,0,0.5}
\newcommand{\CB}[1]{\textcolor{typicalblue}{#1}}

\definecolor{outlierred}{rgb}{0.5,0,0}
\newcommand{\CR}[1]{\textcolor{outlierred}{#1}}

\newcommand{\ifalexfhack}[2]{#1}

\ifalexfhack{}{\usepackage{caption}}

\usepackage{cite}
\usepackage{graphicx}

\newcommand{\Binom}{\mathop{\mathrm{Binom}}}
\newcommand{\email}[1]{{\tt #1}}

\ifalexfhack{}{\DeclareCaptionLabelFormat{AlwaysFigure}{Figure~#2}
\captionsetup{labelformat=AlwaysFigure}}

\title{Your Two Weeks of Fame and Your Grandmother's\footnote{This
    version supercedes the short version of this paper published in
    the proceedings of WWW 2012.}}
\author{James Cook \\
  UC Berkeley%
  \thanks{Work done while interning at Google.}\\
  \email{jcook@cs.berkeley.edu}
  \and
  Atish Das Sarma \\
  eBay Research Labs%
  \thanks{Work done while at Google Research} \\
  \email{atish.dassarma@gmail.com}
  \and
  Alex Fabrikant \\
  Google Research\\
  \email{fabrikant@google.com}
  \and
  Andrew Tomkins \\
  Google Research\\
  \email{atomkins@gmail.com}
}

\begin{document}

\clubpenalty=10000 
\widowpenalty = 10000

\maketitle

\begin{abstract}
Did celebrity last longer in 1929, 1992 or 2009? We investigate the
phenomenon of fame by mining a collection of news articles that spans
the twentieth century, and also perform a side study on a collection
of blog posts from the last 10 years. By analyzing mentions of
personal names, we measure each person's time in the spotlight, using
two simple metrics that evaluate, roughly, the duration of a single
news story about a person, and the overall duration of public interest
in a person. We watched the distribution evolve from 1895 to 2010,
expecting to find significantly shortening fame durations, per the
much popularly bemoaned shortening of society's attention spans and
quickening of media's news cycles. Instead, we conclusively
demonstrate that, through many decades of rapid technological and
societal change, through the appearance of Twitter, communication
satellites, and the Internet, fame durations did not decrease, neither
for the typical case nor for the extremely famous, with the last
statistically significant fame duration decreases coming in the early
20th century, perhaps from the spread of telegraphy and telephony.
Furthermore, while median fame durations stayed persistently constant,
for the most famous of the famous, as measured by either volume or
duration of media attention, fame durations have actually trended
gently upward since the 1940s, with statistically significant
increases on 40-year timescales. Similar studies have been done with
much shorter timescales specifically in the context of information
spreading on Twitter and similar social networking sites. To
the best of our knowledge, this is the first massive scale study of
this nature that spans over a century of archived data, thereby
allowing us to track changes across decades.

\vspace{0.5\baselineskip}\noindent \textit{Keywords}: culturomics,
media, attention modeling, social media, time series, historical
trends, fame duration, news archives
\end{abstract}

\section{Introduction}

Beginning in the 19th century, long-distance communication
transitioned from foot to telegraph on land, and from sail to steam to
cable by sea.  Each new form of technology began with a limited number
of dedicated routes, then expanded to reach a large fraction of the
accessible audience, eventually resulting in near-complete deployment
of digital electronic communication.  Each transition represented an
opportunity for news to travel faster, break more uniformly, and reach
a broad audience closer to its time of inception.

Even today, the increasing speed of the news cycle is a common theme
in discussions of the societal implications of technology.  Stories
break faster, are covered in less detail, and news sources quickly
move on to other topics.  Online and cable outlets aggressively search
for novelty in order to keep eyeballs glued to screens.  Popular
non-fiction dedicates significant coverage to this trend, which by
2007 prompted \textit{The Onion}\footnote{{\tt
    http://www.theonion.com/}}, a satirical website, to offer the
following commentary on cable news provider CNN's\footnote{{\tt
    http://www.cnn.com/}} offerings\cite{OnionCNN}: ``CNN is widely
credited with initiating the acceleration of the modern news cycle
with the fall 2006 debut of its spin-off channel CNN:24, which
provides a breaking news story, an update on that story, and a news
recap all within 24 seconds.''

With this speed-up of the news cycle comes an associated concern that,
whether or not causality is at play, attention spans are shorter, and
consumers are only able to focus for progressively briefer periods on
any one news subject.  Stories that might previously have occupied
several days of popular attention might emerge, run their course, and
vanish in a single day.  This popular theory is consistent with a
suggestion by Herbert Simon~\cite{simon1971designing} that as the
world grows rich in information, the attention necessary to process
that information becomes a scarce and valuable resource.

The speed of the news cycle is a difficult concept to pin down.  We
focus our study on the most common object of news: the individual.  An
individual's fame on a particular day might be thought of as the
probability with which a reader picking up a news article at random
would see their name.  From this idea we develop two notions of the
duration of the interval when an individual is in the news.  The first
is based on fall-off from a peak, and intends to capture the spike
around a concrete, narrowly-defined news story.  The second looks for
period of sustained public interest in an individual, from the time
the public first notices that person's existence until the public
loses interest and the name stops appearing in the news.  We study the
interaction of these two notions of ``duration of fame'' with the
radical shifts in the news cycle we outline above.  For this purpose,
we employ Google's public news archive corpus, which contains over
sixty million pages covering 250 years, and we perform what we believe
to be the first study of the dynamics of fame over such a time period.

Data within the archive is heterogeneous in nature, ranging from
directly captured digital content to optical character recognition
employed against microfilm representations of old newspapers.  The
crawl is not complete, and we do not have full information about which
items are missing.  Rather than attempt topic detection and tracking
in this error-prone environment, we instead directly employ a
recognizer for person names to all content within the corpus; this
approach is more robust, and more aligned with our goal of studying
fame of individuals.

Based on these different notions of periods of reference to a particular
person, we develop at each point in time a distribution over the duration of
fame of different individuals.

Our expectation upon undertaking this study was that in early periods,
improvements to communication would cause the distribution of duration
of coverage of a particular person to shrink over time. We
hypothesized that, through the 20th century, the continued deployment
of technology, and the changes to modern journalism resulting from
competition to offer more news faster, would result in a continuous
shrinking of fame durations, over the course of the century into the
present day.

\noindent\emph{Summary of findings.}

We did indeed observe fame durations shortening somewhat in the early
20th century, in line with our hypothesis regarding accelerating
communications. However, from 1940 to 2010, we see quite a different
picture.  Over the course of 70 years, through a world war, a global
depression, a two order of magnitude growth in (available) media
volume, and a technological curve moving from party-line telephones to
satellites and Twitter, both of our fame duration metrics showed that
\textbf{neither the typical person in the news, i.e. the median fame
  duration, nor the most famous, i.e. high-volume or long-duration
  outliers, experienced any statistically significant decrease in fame
  durations}.

As a matter of fact, \textbf{the bulk of the distribution, as
  characterized by median fame durations, stayed constant throughout
  the entire century-long span of the news study and was also
  the same through the decade of Blogger posts on which we ran
  the same experiments}. As another heuristic characterization of the
bulk of the distribution, both news and Blogger data produced
roughly comparable parameters when fitted to a power law: an exponent
of around -2.5, although with substantial error bars, suggesting
that the fits were mediocre.

Furthermore, when we focused our attention on the very famous, by
various definitions, all signs pointed to a slow but observable growth
in fame durations. \textbf{From 1940 onward, on the scale of 40-year
  intervals, we found statistically significant fame duration growth
  for the ``very famous''}, defined as either:
\begin{itemize}
\item people whose fame lasts exceptionally long:
90th and 99th percentiles of fame duration distributions; or
\item exceptionally highly-discussed people: using distributions among
just the top 1000 people or the top 0.1\% of people by number of
mentions within each year.
\end{itemize}

In the case of taking the 1000 most-often-mentioned names in each
year, the increasing could be explained as follows: as the corpus
increases in volume toward later years, a larger number of names
appear, representing more draws from the same underlying distribution
of fame durations.  The quantiles of the distribution of duration for
the top 1000 elements will therefore grow over time as the corpus
volume increases.  On the other hand, our experiments that took the
top 0.1\% most-often-mentioned names, or the top quantiles of
duration, still showed an increasing trend.  We therefore conclude
that the increasing trend is not completely caused by an increase in
corpus volume.

To summarize, we find that the most famous figures in today's news
stay in the limelight for longer than their counterparts did in the
past.  At the same time, however, the average newsworthy person
remains in the limelight for essentially the same amount of time today
as in the past.


\section{Working with the news corpus}

\begin{figure}
  \includegraphics[width=\columnwidth]{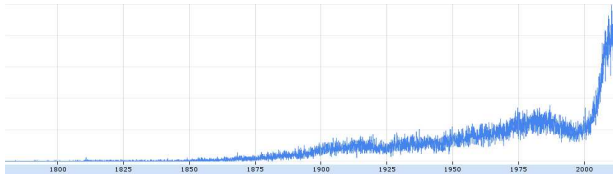}
  \caption{
    \label{fig:article_volume}
    The volume of news articles by date.}
\end{figure}

We perform our main study on a collection of the more than 60 million
news articles in the Google archive that are both (1) in English, and
(2) searchable and readable by Google News users at no cost.  In
Section~\ref{sec:BloggerResults}, we cross-validate our observations against
the corpus of public blog posts on Blogger, which is described there.

The articles of the news corpus span a wide range of time, with the
relative daily volume of articles over the range of the corpus shown
in Figure~\ref{fig:article_volume}. There are a handful of articles
from the late 18th century onward, and the article coverage grows rapidly
over the course of the 19th century. From the last decade of the 19th
century through the end of the corpus (March 2011), there is
consistently a very substantial volume of articles per day, as well as
a wide diversity of publications. For the sake of statistical
significance, our study focuses on the years 1895--2011.

The news corpus contains a mix of modern articles obtained from the
publisher in the original digital form, as well as historical articles
scanned from archival microform and OCRed, both by Google and by third
parties. For scanned articles, per-article metadata such as titles,
issue dates, and boundaries between articles are also derived
algorithmically from the OCRed data, rather than manually curated.

Our study design was driven by several features that we discovered in
this massive corpus. We list them here to explain our study
design. Also, data mining for high-level behavioral patterns in a
diachronous, heterogeneous, partially-OCRed corpus of this scale is
quite new, precedented on this scale perhaps only by \cite{ngrams}
(which brands this new area as ``culturomics''). But, with the rapid
digitization of historical data, we expect such work to boom in the
near future. We thus hope that the lessons we have learned about this
corpus will also be of independent interest to others examining this
corpus and other similar archive corpuses.

\subsection{Corpus features, misfeatures, and missteps}

\subsubsection{News mentions as a unit of attention}
\label{sec:articles}

Our 116-year study of the news corpus aims to extend the rich
literature studying topic attention in online social media like
Twitter, typically over the span of the last 3--5 years. Needless to
say, 100-year-old printed newspapers are an imperfect proxy for the
attention of individuals, which has only recently become directly
observable via online behavior. Implicit in the heart of our study is
the assumption that news articles are published to serve an audience,
and the media makes an effort, even if imperfect, to cater to the
audience's information appetites. We coarsely approximate a unit of
attention as one occurrence in a Google News archive article, and we leave open a
number of natural extensions to this work, such as weighting articles
by historical publication subscriber counts, or by size and position
on the printed page.

Due to the automated OCR process, not every ``item'' in the
corpus can be reasonably declared a news article.
For example, a single photo caption might be extracted as an independent
article, or a sequence of articles on the same page might be
misinterpreted as a single article.
Rather than weighting each of these corpus items equally when measuring the
attention paid to a name, we elected to count multiple mentions of a name
within an item separately, so that articles will tend to count more than
captions, and there is no harm in mistakenly grouping multiple articles as one.

We manually
examined (A) a uniform sample of 50 articles from the whole corpus
(which, per Fig.~\ref{fig:article_volume}, contains overwhelmingly
articles from the last decade), and (B) a uniform sample of 50
articles from 1900--1925. We classified each sample into:
\vspace{-5pt}
\begin{itemize}
\item News articles: timely content, formatted as a stand-alone ``item'',
published without external sponsorship, for the benefit of
part of the publication's audience,
\item News-like items: non-article text chunks where a name mention
  can qualify as that person being ``in the news'' --- e.g. photo
  captions or inset quotes,
\item Non-news: ads and paid content, sports scores, recipes, news
  website comments miscategorized as news, etc.
\end{itemize}
\vspace{-5pt}
The number of items of each type in the two samples are given in the following
table.
\smallskip \\
\begin{tabular}{r|c|c}
    & \scriptsize full corpus sample & \scriptsize 1900--1925 sample \\
    \hline
    news articles & 31 & 28 \\
    news-like items & 3 & 2 \\
    non-news items & 16 & 20
\end{tabular} \ \\
We expect that the similarity in these
distributions should result in minimal noise in the cross-temporal
comparisons, and leave to future work the task of automatically
distinguishing real news stories from non-news.

\subsubsection{Compensating for coverage}
\label{sec:corpuscoverage}

Even once we discard the more sparsely covered 18th and 19th centuries,
there is still more than an order of magnitude difference between
article volume in 1895 and 2011. We address these coverage differences
by downsampling the data down to the same number of articles for each
month in this range. We address the nuanced effects of this
downsampling on our methodology in Section~\ref{sec:Subsampling}.

\subsubsection{Evolution of discourse and media --- why names?}
\label{whynames}

We set out originally to understand changes in the public's attention
as measured by news story topics. There are a myriad heuristics to
define a computationally feasible model of a ``single topic'' that can
be thought to receive and lose the public's attention. But over the
course of a century, the changes in society, media formatting,
subjects of public discourse, writing styles, and even language itself
are substantial enough that neither sophisticated statistical models
trained on plentiful, well-curated training data from modern media nor
simple generic approaches like word co-occurrence in titles are
guaranteed to work well. Very few patterns connect articles from 1910
newspapers' ``social'' sections (now all but forgotten) about tea at
Mrs.\ Smith's, to 1930 articles about the arrival of a trans-oceanic
liner, to 2009 articles about a viral Youtube video.

\begin{figure}
  \includegraphics[width=0.5\textwidth]{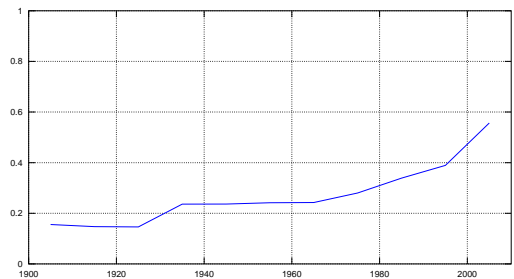}
  \caption{
    \label{fig:havenames}Articles with recognized personal names per decade
  }
\end{figure}

After trying out general proper noun phrases produced inconclusively
noisy results, we decided to focus on occurrences of personal names,
detected in the text by a proprietary state-of-the-art statistical
recognizer. Personal names have a relatively stable presence in the
media: even with high OCR error rates in old microform, over 1/7th of
the articles even in the earliest decades since 1900 contain
recognized personal names (see Figure~\ref{fig:havenames}).

But personal names are not without historical caveats, either.  A
woman appearing in 2005 stories as ``Jane Smith'' would be much more
likely to be exclusively referenced as ``Mrs.\ Smith'', or even
``Mrs.\ John Smith'', in 1915. Also, the English-speaking world was
much more Anglo-centric in 1900 than now, with much less diversity of
names. An informal sample suggests that most names with non-trivial
news presence 100 years ago referred overwhelmingly to a single bearer
of that name for the duration of a particular news topic, but many
names are not unique when taken across the duration of the whole
corpus --- for instance, ``John Jacob Astor'', appearing in the news
heavily over several decades
(Fig.~\ref{fig:timeline:john_jacob_astor}), in reference to a number
of distinct relatives. On account of both of these phenomena, among
others, we aim to focus on name appearance patterns that are most
likely to represent a single news story or contiguous span of public attention
involving
that person, rather than trying to model the full media ``lifetime''
of individuals, as we had considered doing at the start of this
project.

\subsubsection{OCR errors in data and metadata}
\label{sec:ocr}

We empirically discovered another downfall of studying long-term
``media lifetimes'' of individuals. In an early experiment, we
measured, for each personal name, the 10th and 90th percentiles of the
dates of that name's occurrence in the news. We then looked at the time
interval between 10th and 90th percentiles, postulating that a large
enough fraction of names are unique among newsworthy individuals that
the distribution of these \emph{inter-quantile gaps} could be a robust
measure of media lifetime. After noticing a solid fraction of the
dataset showing inter-quantile gaps on the scale of 10-30 years, we
examined a heat map of gap durations, and discovered a regular pattern
of gap durations at exact-integer year offsets, which, other than for
Santa Claus, Guy Fawkes, and a few other clear exceptions, seemed an
improbable phenomenon.

This turned out to be an artifact of OCRed metadata. In particular,
the culprit was single-digit OCR errors in the \emph{scanned article
  year}. While year errors are relatively rare, every long-tail name
that occurred in fewer than 10 articles (often within a day or two of
each other), and had a mis-OCRed error for one of those occurrences
contributed probability mass to integral-number-of-years media
lifetimes. As extra evidence, the heat map had a distinct outlier
segment of high probability mass for inter-quantile range of exactly
20 years, starting in the 1960s and ending in the 1980s --- the digits 6 and
8 being particularly easy to mistake on blurry microfilm. Note that
short-term phenomena are relatively safe from OCR date errors, thanks
to the common English convention of written-out month names, and to the
low impact of OCR errors in the day number.

OCR errors in the article text itself are ubiquitous. Conveniently,
the edit distance between two recognizable personal names is rarely
very short, so by agreeing to discard any name that occurs only once
in the corpus, we are likely to discard virtually all OCR errors as
well, with no impact on data on substantially newsworthy people. We
should note that OCR errors are noticeably more frequent on older
microfilm, but the reasonable availability of recognizable personal
names even in 100-year-old articles, per Fig.~\ref{fig:havenames},
suggests that this problem is not dire. A manually-coded sample of 50
articles with recognized names from the first decade of the 1900s
showed only 8 out of 50 articles having incorrectly recognized names
(including both OCR errors and non-names mis-tagged as names).

\subsubsection{Simultaneity and publishing cycles}
\label{sec:simultaneity}

There are also pitfalls with examining short timelines. In the
earliest decades we examine, telegraph was widely available to news
publishers, but not fully ubiquitous, with rural papers often
reporting news ``from the wire'' several days after the event. An
informal sample seems to suggest that most news by 1900 propagated
across the world on the scale of a few days. Also, many publications
in the corpus until the last 20 years or so were either published
exclusively weekly or, in the case of Sunday newspaper issues, had
substantially higher volume once a week, resulting in many otherwise
obscure names having multiple news mentions separated by one week ---
a rather different phenomenon than a person remaining in the daily
news for a full week. On account of both of these, we
generally disregard news patterns that are shorter than a few days in
our study design.

\section{Measuring Fame}

\label{sec:Method}

We begin by producing a list of names for each article.  To do this,
we extract short capitalized phrases from the body text of each
article, and keep phrases recognized by an algorithm
to be personal names.

For every name that appears in the input, we consider that name's
\emph{timeline}, which is the multiset of dates at which that name
appears, including multiple occurrences within an article.  
We intend the timeline to approximate the frequency with which a person
browsing the news at random on a given day would encounter that name.
The accuracy of this approximation will depend on the volume of news articles
available.
In order to avoid the possibility that any trends we detect are caused by
variations in this accuracy caused by variations in the volume of the corpus,
we randomly choose an approximately equal number of articles to work with from
each month.  We describe and analyze this process in Section~\ref{sec:Subsampling}.

In
general, our method can be applied to any collection of timelines.  In
Section~\ref{sec:BloggerResults}, we apply it to names extracted from
blog posts.

\subsection{Finding Periods of Fame}

Once we have computed a timeline for each name that appears in the
corpus, we select a time during which we consider that name to have
had its period of fame, using one of the two methods described
below.  In order to compare the phenomenon of fame at different points
in time, we consider the joint distribution of two variables over the
set of names: the \emph{peak date} and the \emph{duration} of the
name's period of fame.  We try the following two methods to compute
a peak date and duration for each timeline.
\begin{itemize}
\item \textbf{Spike method}.  This method intends to capture the spike
  in public attention surrounding a particular news story.
  We divide time into
  one-week intervals and consider the name's rate of occurrence in
  each interval.  The week with the highest rate is considered to be
  the peak date, and the period extends backward and forward in time as
  long as the rate does not drop below one tenth its maximum rate.
  Yang and Leskovec~\cite{yang2011patterns} used a similar method in
  their study of digital media, using a time scale of hours where we
  use weeks.
\item \textbf{Continuity method}.  This method intends to measure the
  duration of public interest in a person.  We define a name's period
  of popularity to be the longest span of time within which there is
  no seven-day period during which it is not mentioned.  The peak date
  falls halfway between the beginning and the end of the period. We
  find, in Section~\ref{sec:ResultsNews}, that durations are short
  compared to the time span of the study, so using any choice of peak
  date between the beginning and end will produce similar
  distributions.
\end{itemize}

\begin{figure}
  \includegraphics[width=\columnwidth]{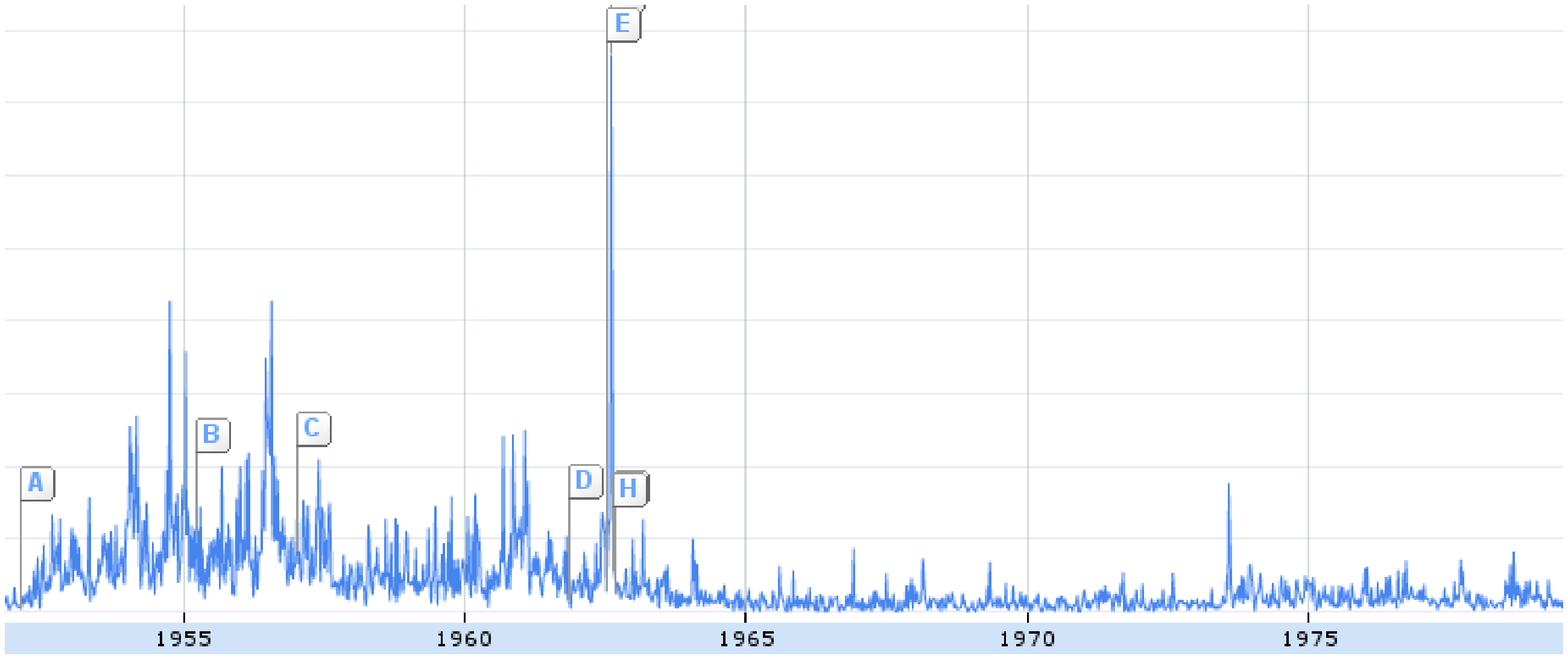} \\
  \includegraphics[width=\columnwidth]{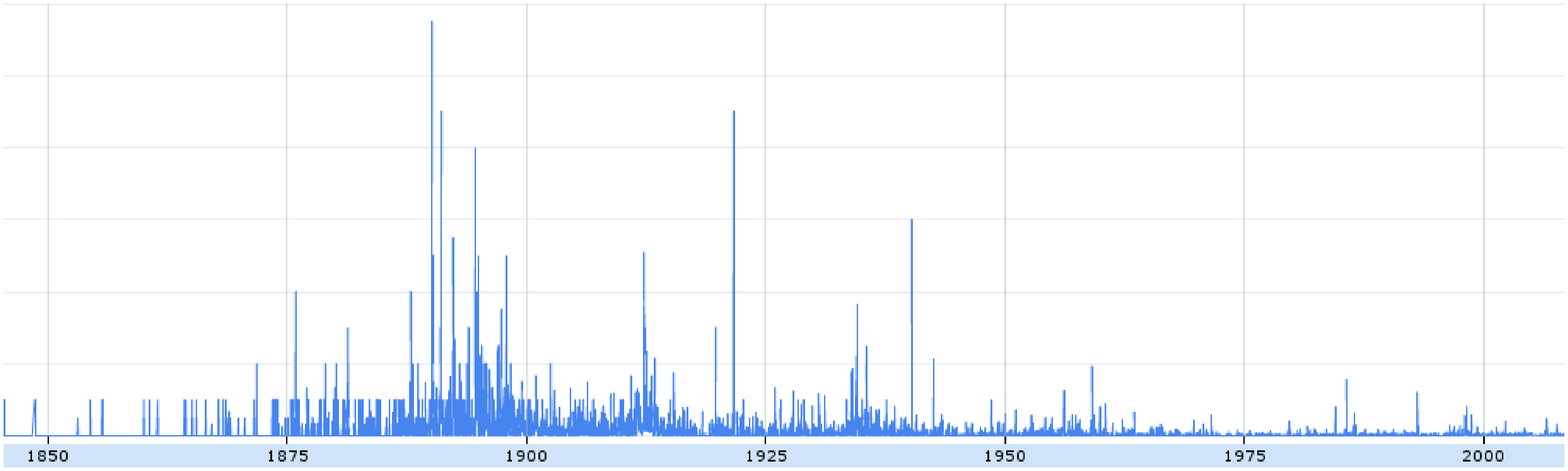}
  \caption{
    \label{fig:timeline:john_jacob_astor}
    Timelines for ``Marilyn Monroe'' (top) and ``John Jacob Astor'' (bot).}
\end{figure}

To demonstrate the distinction between these two methods,
Figure~\ref{fig:timeline:john_jacob_astor} shows the occurrence timeline for
Marilyn Monroe. The ``continuity method'' picks out the bulk of her
fame --- 1952-02-13 (``A'') through 1961-11-15 (``D''), by which point her appearance
in the news was reduced to a fairly low background level. The ``spike
method'' picks out the intense spike in interest surrounding her
death, yielding the range 1962-7-18 (``E'') -- 1962-8-29 (``H'').

Very often these two methods identify short moments of fame
within a much longer context.  For example, in
Figure~\ref{fig:timeline:john_jacob_astor}, we see the timeline for
the name ``John Jacob Astor'', normalized by article counts.  The
spike method identifies as the peak the death of John Jacob
Astor III of the wealthy Astor family, with a duration of 38 days
(March 8 to February 15, 1890).  The continuity method identifies
instead the death of his nephew John Jacob Astor IV, who died on the
Titanic, with a period of five months~\cite{wiki:AstorFamily}.  The
period begins on March 23, 1912, three weeks before the Titanic sank, and
ends August 31.  Many of the later occurrences of the name are
historical mentions of the sinking of the Titanic.

\subsection{Choosing the Set of Names}

\label{sec:FilteringNames}

\paragraph{Basic filtering}
In all our experiments, to reduce noise, we discard the names which occurred
less than ten times, or whose fame durations are less than two days.  (In
both methods, a name whose fame begins Monday and ends Wednesday is
considered to have a duration of two days.)
We also remove
peaks that end in 2011 or later, since these peaks might extend
further if our news corpus extended further in the future.

\paragraph{Top 1000 by year}

For each peak type, we repeat our experiment with the set of names restricted
in the following way.  We counted the total number of times each name appeared
in each year (counting repeats within an article).  For each year, we produced
the set of the 1000 most frequently mentioned names in that year.  We took the
union of these sets over all years, and ran our experiments using only the
names in this set.  Note that a name's peak of popularity need not be the same
year in which that name was in the top 1000: so if a name is included in the
top-1000 set because it was popular in a certain year, we may yet consider that
name's peak date to be a different year.

\paragraph{Top 0.1\% by year}

We consider that filtering to the top 1000 names in each year might introduce
the following undesirable bias.  Suppose names are assigned peak durations
according to some universal distribution, and later years have more names,
perhaps because of the increasing volume of news.  If a name's frequency of
occurrence is proportional to its duration, then selecting the top 1000 names
in each year will tend to produce names with longer durations of fame in years
with a greater number of names.
With this in mind, we considered one more restriction on the set of names.  In
each year $y$, we considered the total number of distinct names $n_y$ mentioned
in that year.  We then collected the top $n_y/1000$ names in each year $y$.
We ran our experiments using only the names in the union of those sets.  As
with the top-1000 filtering, a name's peak date will not necessarily be the
same year for which it was in the top 0.1\% of names.

\subsection{Sampling for Uniform Coverage}

\label{sec:Subsampling}

The spike and continuity methods for identifying periods of fame may
be affected by the volume of articles available in our corpus.  For
example, suppose a name's timeline is generated stochastically, with
every article between February~1 and March~31 containing the name with
a 1\% probability.  If the corpus contains 10000 articles in every
week, then both the spike and continuity methods will
probably decide that the article's duration is two months.  However,
if the corpus contains less than 100 articles in each week, then the
durations will tend to be short, since there will be many weeks during
which the name is not mentioned.

We propose a model for this effect.  Each name \(\nu\) has a ``true'' timeline
which assigns to each day \(t\) a probability \(f_\nu(t)\in[0,1]\) that an
article on that day will mention \(\nu\).%
\footnote{
    In fact, articles could mention the name multiple times, but in the limit
    of a large number of articles, this will not affect our analysis.
}
For each day, there is a total
number of articles \(n_t\); we have no knowledge of the relation between
\(n_t\) and \(\nu\), except that there is some lower bound \(n_t >
n_{\mathrm{min}}\) for all \(t\) within some reasonable range of time.  Then we
suppose the timeline for name \(\nu\) is a sequence of independent random
variables \(X_{\nu,t} \sim \Binom(f_\nu(t), n_t)\).  Our goal is to ensure that
any measurements we take are independent of the values \(n_t\).

To accomplish this independence of news volume, we randomly sampled news
articles so that the expected number in each month was \(n_{\mathrm{min}}\).
Let \(X'_{\nu,t}\) be the number of sampled articles containing name \(\nu\).
If we were to randomly sample \(n_{\mathrm{min}}\) articles without
replacement, then we would have \(X'_{\nu,t} \sim \Binom(f_\nu(t),
n_{\mathrm{min}})\).  Notice that the joint distribution of the random
variables \(X'_{\nu,t}\) is unaffected by the article volumes \(n_t\).  Any
further measurement based on the variables \(X'_{\nu,t}\) will therefore also
be unrelated to the sequence \(n_t\).  In practice, instead of sampling exactly
\(n_{\mathrm{min}}\) articles without replacement, we flipped a biased coin for
each of the \(n_t\) articles at time \(t\), including each article with
probability \(n_{\mathrm{min}}/n_t\).  For a large enough volume of articles,
the resulting measurements will be the same.

We removed all articles published before 1895, since
the months before 1895 had less than our target number \(n_{\mathrm{min}}\) of
articles.
We also removed articles published after the end of the year 2010, to
avoid having a month with news articles at the beginning but not the
end of the month, but with the same number of sampled articles.

As an example of the effect of downsampling, the blue dotted lines in
Figure~\ref{fig:AllArticlesLSPeaks} show the 50th, 90th and 99th
percentiles of the distribution of fame durations using the continuity method.
We see that they
increase suddenly in the last ten years, when our
coverage of articles surges with the digital age.  The red lines show
the same measurement after downsampling: the surge no longer
appears.

\subsection{Graphing the Distributions}
\label{sec:Graphing}

We graph the joint distribution of peak dates and durations in two
different ways.  We consider the set of names which peak in successive
five-year periods.  Among each set of names, we graph the
50th, 90th and 99th percentile
durations of fame.  These appear as darker lines in the graphs; for
example, the top of Fig.~\ref{fig:AllArticlesSpikeyPeaks}
shows the distribution for the spike method.  The lighter
solid red lines show the same three quantiles for shorter three-month
periods.  For comparison, the dashed light blue lines show the same
results if the article sampling described in Sec.~\ref{sec:Subsampling} is
not performed (and articles before 1895 and after 2010 are not
removed).  Fig.~\ref{fig:AllArticlesLSPeaks} shows the same set of
lines using the continuity method.  All the later figures are produced in the
same way,
except they do not include the non-sampled full distributions.

The second type of graph focuses on one five-year period at a time.
The bottom of Fig.~\ref{fig:AllArticlesSpikeyPeaks} shows
a cumulative plot showing the number of names with duration greater
than that shown on the \(x\)-axis.  This is plotted for
many five-year periods.  The graphs of measurements using the
spike method look more like step functions because that method
measures durations in seven-day increments, whereas
the longest-stretch method can yield any number of days.  (Recall that
peaks that last less than two days are removed.)

\subsection{Estimating Power Law Exponents}

We test the hypothesis that the tail of the distribution of fame
durations follows a power law.  For a given five-year period, we
collect all names which peak in that period, and consider 20\% of the
names with the longest fame durations -- that is, we set \(d_{\mathrm{min}}\)
to be the 80th percentile of durations, and consider durations \(d > d_{\mathrm{min}}\).
Among those 20\%, we compute a
maximum likelihood estimate of the power law exponent \(\hat\alpha\),
predicting that the probability of a duration \(d > d_{\mathrm{min}}\) is \(p(d) \propto d^{\hat\alpha}\).
Clauset et al~\cite{journals/siamrev/ClausetSN09} show that the maximum
likelihood estimate \(\hat\alpha\) is given by \(\hat\alpha = 1 +
(\sum_{i=1}^n\ln(d_i/d_{\mathrm{min}}))\).
We include a line on each plot
of cumulative distributions of fame durations, of slope
\(\hat\alpha+1\) on the log-log graph because we plot cumulative
distributions rather than density functions.  The \(\hat\alpha\)
values we measure are discussed in the following sections, and
summarized in Figure~\ref{fig:ArticlePowerLawExponents} for the news
corpus and Figure~\ref{fig:BloggerPowerLawExponents} for the blog
corpus.

\subsection{Statistical Measurements}
\label{sec:StatMeas}

We used bootstrapping to estimate the uncertainty in the four statistics we
measured: the 50th, 90th and 99th percentile durations and of the best-fit
power law exponents.  For selected five-year periods, we sampled \(|S|\) names
with replacement from the set \(S\) of names that peaked in that period of
time.  For each statistic, we repeated this process 25000 times, and reported
the range of numbers within which 99\% of our samples fell.  The results are
presented in Figures \ref{fig:ArticlePowerLawExponents} (for the news corpus)
and \ref{fig:BloggerPowerLawExponents} (for the blog corpus).


\begin{figure*}[t]
  {\small \begin{tabular}{c|c|c|cccc}
method & filtering & period & 50th \%ile (days) & 90th \%ile (days) & 99th \%ile (days) & power law exponent \\
\hline
spike & all & 1905-9
& \CB{7 (7 .. 7)}
& \CR{28 (28 .. 28)}
& \CR{91 (78 .. 106)}
& -2.45 (-2.55 .. -2.21)
\\
spike & all & 1925-9
& \CB{7 (7 .. 7)}
& \CR{28 (28 .. 28)}
& \CR{65 (63 .. 78)}
& -2.63 (-2.74 .. -2.33)
\\
spike & all & 1945-9
& \CB{7 (7 .. 7)}
& \CR{21 (21 .. 28)}
& \CR{56 (49 .. 63)}
& -2.44 (-2.50 .. -2.38)
\\
spike & all & 1965-9
& \CB{7 (7 .. 7)}
& \CR{21 (21 .. 28)}
& \CR{63 (56 .. 70)}
& -2.37 (-2.44 .. -2.31)
\\
spike & all & 1985-9
& \CB{7 (7 .. 7)}
& \CR{21 (21 .. 28)}
& \CR{70 (63 .. 78)}
& -2.32 (-2.36 .. -2.27)
\\
spike & all & 2005-9
& \CB{7 (7 .. 7)}
& \CR{28 (28 .. 28)}
& \CR{84 (78 .. 91)}
& -2.48 (-2.53 .. -2.43)
\\
\hline
spike & top 1000 & 1905-9
& \CR{21 (21 .. 21)}
& 63 (56 .. 70)
& 155 (133 .. 192)
& -2.75 (-3.15 .. -2.56)
\\
spike & top 1000 & 1925-9
& \CR{21 (14 .. 21)}
& 49 (46 .. 56)
& 91 (78 .. 113)
& -3.22 (-3.74 .. -2.99)
\\
spike & top 1000 & 1945-9
& \CR{21 (14 .. 21)}
& 49 (42 .. 49)
& 91 (70 .. 130)
& -3.33 (-3.73 .. -2.89)
\\
spike & top 1000 & 1965-9
& \CR{21 (21 .. 21)}
& 56 (49 .. 63)
& 119 (99 .. 164)
& -2.90 (-3.54 .. -2.65)
\\
spike & top 1000 & 1985-9
& \CR{21 (21 .. 28)}
& 63 (56 .. 78)
& 161 (121 .. 366)
& -2.85 (-3.19 .. -2.57)
\\
spike & top 1000 & 2005-9
& \CR{35 (28 .. 35)}
& 99 (84 .. 119)
& 309 (224 .. 439)
& -2.64 (-2.96 .. -2.44)
\\
\hline
spike & top 0.1\% & 1905-9
& \CR{35 (28 .. 42)}
& 122 (91 .. 155)
& 289 (161 .. 381)
& -2.82 (-3.96 .. -2.36)
\\
spike & top 0.1\% & 1925-9
& \CR{28 (21 .. 35)}
& 63 (56 .. 82)
& 145 (91 .. 218)
& -3.49 (-4.82 .. -2.92)
\\
spike & top 0.1\% & 1945-9
& \CR{21 (21 .. 28)}
& 56 (49 .. 67)
& 133 (84 .. 161)
& -3.35 (-4.32 .. -2.78)
\\
spike & top 0.1\% & 1965-9
& \CR{28 (21 .. 35)}
& 70 (63 .. 99)
& 162 (119 .. 494)
& -2.90 (-3.77 .. -2.47)
\\
spike & top 0.1\% & 1985-9
& \CR{35 (28 .. 35)}
& 90 (70 .. 113)
& 327 (140 .. 443)
& -2.66 (-3.13 .. -2.35)
\\
spike & top 0.1\% & 2005-9
& \CR{35 (35 .. 42)}
& 119 (99 .. 140)
& 338 (263 .. 557)
& -2.76 (-3.10 .. -2.44)
\\
\hline
continuity & all & 1905-9
& \CB{7 (7 .. 7)}
& \CR{20 (19 .. 21)}
& \CR{70 (64 .. 79)}
& -2.67 (-2.76 .. -2.59)
\\
continuity & all & 1925-9
& \CB{7 (7 .. 7)}
& \CR{18 (17 .. 19)}
& \CR{64 (56 .. 71)}
& -2.64 (-2.72 .. -2.53)
\\
continuity & all & 1945-9
& \CB{7 (7 .. 7)}
& \CR{16 (15 .. 16)}
& \CR{53 (49 .. 58)}
& -2.74 (-2.82 .. -2.66)
\\
continuity & all & 1965-9
& \CB{7 (7 .. 7)}
& \CR{17 (16 .. 18)}
& \CR{66 (58 .. 75)}
& -2.58 (-2.69 .. -2.52)
\\
continuity & all & 1985-9
& \CB{7 (7 .. 7)}
& \CR{18 (17 .. 18)}
& \CR{77 (71 .. 83)}
& -2.48 (-2.56 .. -2.44)
\\
continuity & all & 2005-9
& \CB{7 (7 .. 7)}
& \CR{21 (20 .. 21)}
& \CR{101 (96 .. 108)}
& -2.43 (-2.46 .. -2.40)
\\
\hline
continuity & top 1000 & 1905-9
& \CR{24 (23 .. 26)}
& 69 (62 .. 76)
& 166 (136 .. 229)
& -3.01 (-3.35 .. -2.70)
\\
continuity & top 1000 & 1925-9
& \CR{22 (21 .. 24)}
& 58 (53 .. 66)
& 176 (131 .. 338)
& -3.01 (-3.39 .. -2.67)
\\
continuity & top 1000 & 1945-9
& \CR{27 (25 .. 29)}
& 66 (57 .. 80)
& 211 (169 .. 332)
& -2.92 (-3.32 .. -2.59)
\\
continuity & top 1000 & 1965-9
& \CR{34 (32 .. 35)}
& 92 (81 .. 104)
& 262 (203 .. 622)
& -2.75 (-3.11 .. -2.48)
\\
continuity & top 1000 & 1985-9
& \CR{52 (49 .. 56)}
& 135 (118 .. 147)
& 312 (231 .. 739)
& -3.20 (-3.62 .. -2.83)
\\
continuity & top 1000 & 2005-9
& \CR{87 (80 .. 91)}
& 229 (211 .. 250)
& 649 (532 .. 752)
& -2.97 (-3.32 .. -2.75)
\\
\hline
continuity & top 0.1\% & 1905-9
& \CR{66 (59 .. 79)}
& 146 (126 .. 176)
& 968 (209 .. 4287)
& -3.29 (-5.20 .. -2.24)
\\
continuity & top 0.1\% & 1925-9
& \CR{53 (47 .. 61)}
& 125 (104 .. 161)
& 476 (258 .. 2498)
& -2.67 (-3.72 .. -2.20)
\\
continuity & top 0.1\% & 1945-9
& \CR{57 (52 .. 66)}
& 150 (123 .. 194)
& 419 (218 .. 1089)
& -3.19 (-4.26 .. -2.52)
\\
continuity & top 0.1\% & 1965-9
& \CR{69 (61 .. 79)}
& 168 (143 .. 214)
& 713 (261 .. 874)
& -3.01 (-4.01 .. -2.45)
\\
continuity & top 0.1\% & 1985-9
& \CR{85 (78 .. 94)}
& 187 (158 .. 216)
& 732 (276 .. 892)
& -3.40 (-4.30 .. -2.80)
\\
continuity & top 0.1\% & 2005-9
& \CR{113 (107 .. 119)}
& 271 (246 .. 306)
& 681 (614 .. 874)
& -3.16 (-3.59 .. -2.85)
\\
\end{tabular}
}
  \caption{
    \label{fig:ArticlePowerLawExponents}
    Percentiles and best-fit power-law exponents for five-year periods
    of the news corpus.  Each entry shows the estimate based on the
    corpus, and the 99\% boostrap interval in parentheses, as
    described in Section~\ref{sec:StatMeas}. Results discussed in
    section~\ref{sec:ResultsNews}.}
\end{figure*}
\begin{figure*}[t]
  {\small \begin{tabular}{c|c|c|cccc}
method & filtering & period & 50th \%ile (days) & 90th \%ile (days) & 99th \%ile (days) & power law exponent \\
\hline
spike & all & 2000-4
& 7 (7 .. 7)
& 35 (28 .. 35)
& 123 (84 .. 189)
& -2.37 (-2.52 .. -2.23)
\\
spike & all & 2005-9
& 7 (7 .. 7)
& 28 (21 .. 28)
& 75 (63 .. 84)
& -2.34 (-2.76 .. -2.27)
\\
\hline
spike & top 1000 & 2000-4
& 21 (14 .. 21)
& 56 (49 .. 63)
& 265 (148 .. 479)
& -2.51 (-2.83 .. -2.18)
\\
spike & top 1000 & 2005-9
& 14 (14 .. 21)
& 49 (42 .. 54)
& 109 (91 .. 151)
& -2.74 (-3.03 .. -2.41)
\\
\hline
spike & top 0.1\% & 2000-4
& 39 (28 .. 56)
& 189 (106 .. 305)
& 717 (286 .. 840)
& -2.26 (-3.05 .. -1.85)
\\
spike & top 0.1\% & 2005-9
& 28 (25 .. 35)
& 88 (74 .. 102)
& 213 (113 .. 1674)
& -3.29 (-5.40 .. -2.23)
\\
\hline
continuity & all & 2000-4
& 7 (7 .. 7)
& 22 (20 .. 23)
& 114 (95 .. 160)
& -2.38 (-2.49 .. -2.28)
\\
continuity & all & 2005-9
& 6 (6 .. 7)
& 18 (17 .. 19)
& 80 (66 .. 93)
& -2.62 (-2.72 .. -2.53)
\\
\hline
continuity & top 1000 & 2000-4
& 20 (18 .. 21)
& 71 (59 .. 83)
& 387 (237 .. 819)
& -2.32 (-2.54 .. -2.12)
\\
continuity & top 1000 & 2005-9
& 21 (20 .. 22)
& 59 (53 .. 73)
& 408 (211 .. 1057)
& -2.37 (-2.62 .. -2.18)
\\
\hline
continuity & top 0.1\% & 2000-4
& 102 (89 .. 123)
& 372 (236 .. 768)
& 2010 (768 .. 2238)
& -2.24 (-3.15 .. -1.86)
\\
continuity & top 0.1\% & 2005-9
& 83 (70 .. 93)
& 302 (193 .. 617)
& 2083 (954 .. 2991)
& -2.12 (-2.75 .. -1.79)
\\
\end{tabular}
}
  \caption{
    \label{fig:BloggerPowerLawExponents}
    Percentiles and best-fit power-law exponents for five-year periods
    of the blog corpus. Each entry shows the estimate based on the
    corpus, and the 99\% boostrap interval in parentheses, as
    described in Section~\ref{sec:StatMeas}. Results discussed in
    Section~\ref{sec:BloggerResults}.}
\end{figure*}

\begin{figure}[t]
  \includegraphics{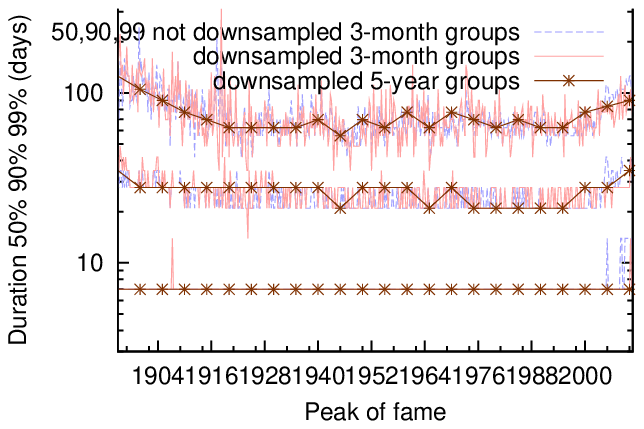}
  \includegraphics{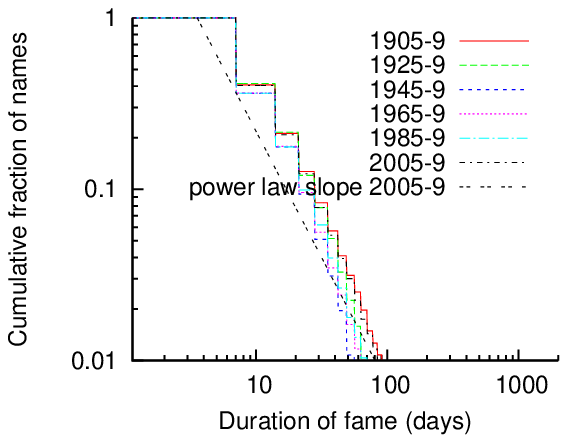}
  \caption{
    \label{fig:AllArticlesSpikeyPeaks}
    Fame durations measured using the spike method, plotted
    as the 50th, 90th and 99th percentiles over time (top)
    and for specific five-year periods (bottom).
    The bottom graph also includes a line showing the
    max-likelihood power law exponent for the years 2005-9.  (The
    slope on the graph is one plus the exponent from
    Fig.~\ref{fig:ArticlePowerLawExponents}, since we graph
    the cumulative distribution function.)  To illustrate the effect of
    sampling for uniform article volume, the first graph includes measurements
    taken before sampling; see Sec.~\ref{sec:Subsampling}.
    Section~\ref{sec:Graphing} describes the format of the graphs in detail.}
\end{figure}

\begin{figure}[t]
  \includegraphics{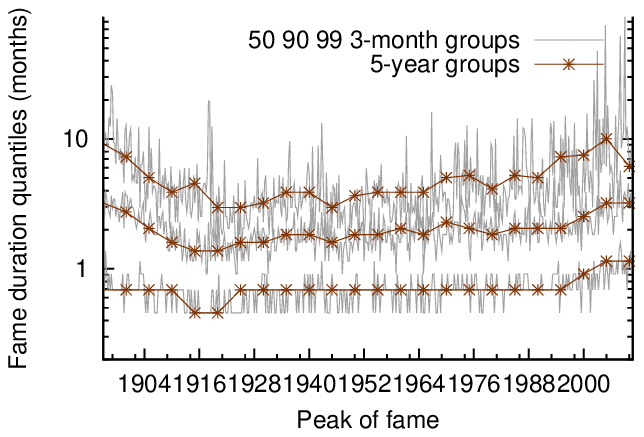}
  \includegraphics{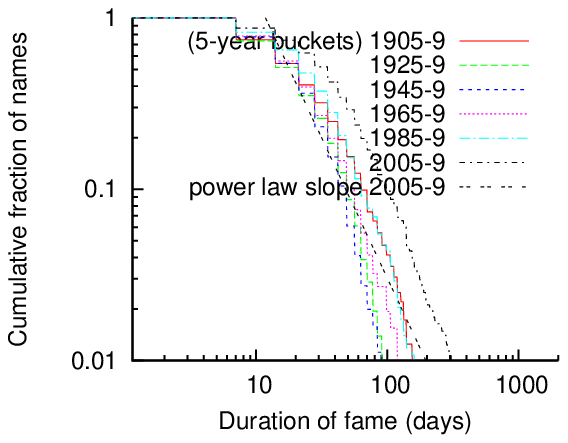}
  \caption{
    \label{fig:DsArticlesConstTeSpikeyPeaks}
    Fame durations, restricting to the union of the 1000
    most-mentioned names in every year, using the spike
    method to identify periods of fame.}
\end{figure}

\begin{figure}[t]
  \includegraphics{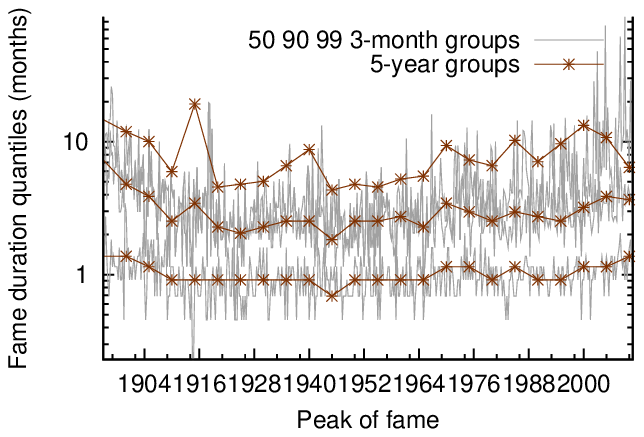}
  \includegraphics{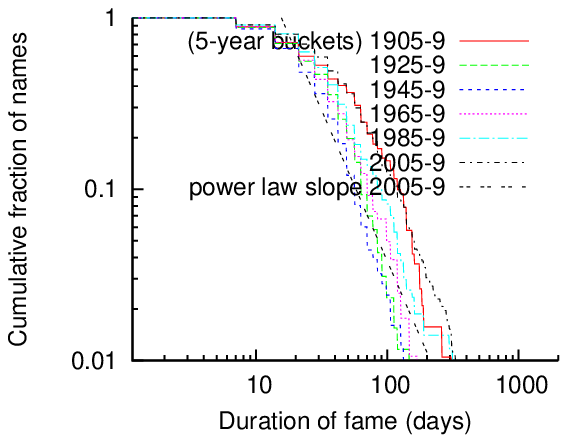}
  \caption{
    \label{fig:DsArticlesFracTeSpikeyPeaks}
    Fame durations, restricting to the union of the 0.1\%
    most-mentioned names in every year, measured using the spike
    method.}
\end{figure}

\begin{figure}
  \includegraphics{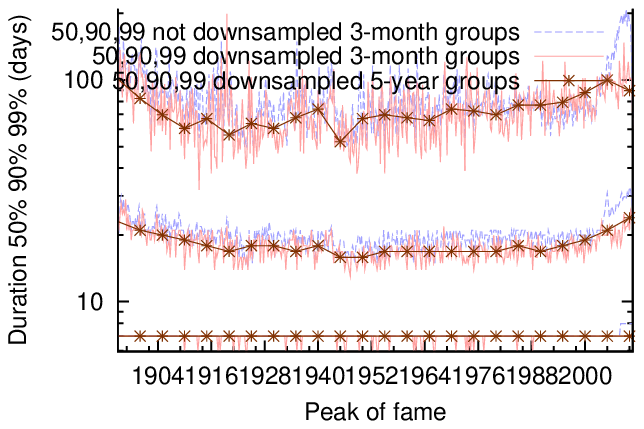}
  \includegraphics{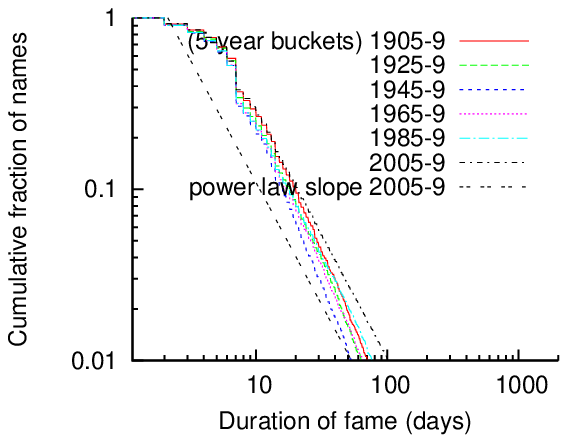}
  \caption{
    \label{fig:AllArticlesLSPeaks}
    Fame durations measured using the continuity method, plotted
    as the 50th, 90th and 99th percentiles over
    time (top), and for specific five-year
    periods (bottom).  To illustrate the effect of sampling, the first graph
    includes measurements taken before sampling; see
    Section~\ref{sec:Subsampling}.
    Section~\ref{sec:Graphing} describes the format of the graphs in detail.
  }
\end{figure}

\begin{figure}[t]
  \includegraphics{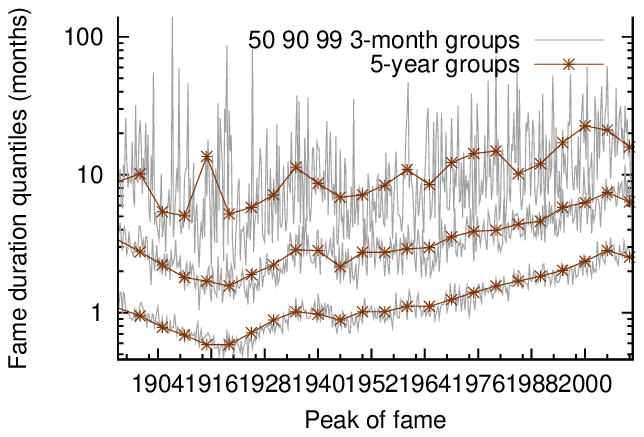}
  \includegraphics{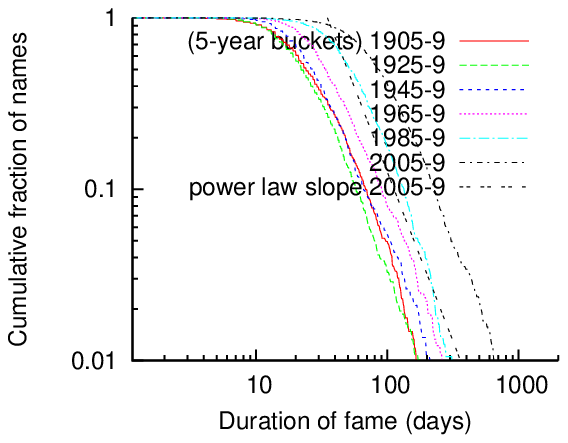}
  \caption{
    \label{fig:DsArticlesConstTeLSPeaks}
    Fame durations, restricting to the union of the 1000
    most-mentioned names in every year, measured using the continuity
    method.}
\end{figure}

\begin{figure}
  \includegraphics{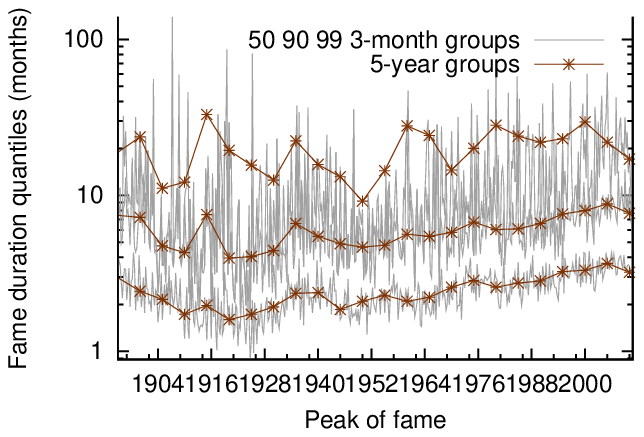}
  \includegraphics{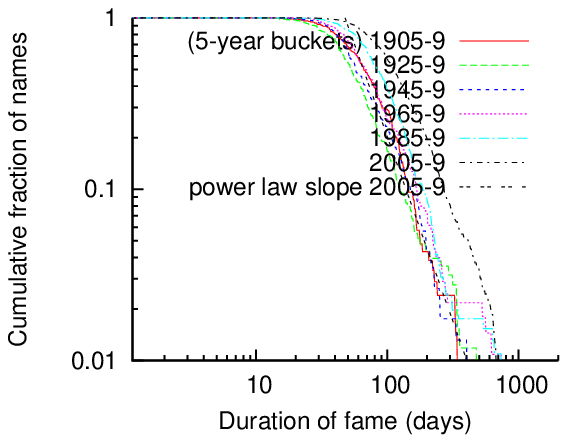}
  \caption{
    \label{fig:DsArticlesFracTeLSPeaks}
    Fame durations, restricting to the union of the 0.1\%
    most-mentioned names in every year, measured using the continuity
    method.}
\end{figure}

\section{Results: News Corpus}
\label{sec:ResultsNews}

We measure periods of popularity using the spike and continuity methods
described in Section~\ref{sec:Method}, and in each case plot the distribution
of duration as it changes over time.

Figures \ref{fig:AllArticlesSpikeyPeaks} and~\ref{fig:AllArticlesLSPeaks} show
the evolution of the distribution of fame durations for the full set of names
in the corpus (after the basic filtering described in
Section~\ref{sec:FilteringNames}) using the spike and continuity methods,
respectively.  (Section~\ref{sec:Graphing} describes the format of the graphs
in detail.)

\paragraph{Median durations}
For the entire period we studied, the median fame duration did not
decrease, as we had expected, but rather remained completely constant
at exactly 7 days, for both the spike and the continuity peak
measurement methods. For the spike method alone, this would not have
been surprising. Peaks measured by the spike method are discretized to
multiples of weeks, so a perennial median of 7 days just shows that
multi-week durations have never been common. On the other hand, the
continuity method freely admits fame durations in increments of 1 day,
with only 1-day-long peaks filtered out. Yet, the median has remained
at exactly 7 days for all the years studied, and, per the full-corpus
``50th percentile'' measurements, shown in blue in
Figure~\ref{fig:ArticlePowerLawExponents}, for all decades where we've
tried bootstrapping, 99\% of bootstrapped samples also matched the
7-day measurement exactly (for the continuity method and, less
surprisingly, for the spike method). This gives strong statistical
significance to the claim that 7 days is indeed a very robust
measurement of typical fame duration, which has not varied in a
century.

\paragraph{The most famous}

We next consider specially the fame durations of the most famous
names, in two correlated, but distinct senses of ``most famous'':
\begin{itemize}
\item ``Duration outliers'' --- people whose \textbf{fame lasts much longer than
  typical}, as measured by the 90th and 99th percentiles of fame
  durations within each year.  These correspond to the top two lines
  in the timelines of Figures \ref{fig:AllArticlesSpikeyPeaks}
  and~\ref{fig:AllArticlesLSPeaks}, and the columns ``90 \%ile'' and
  ``99 \%ile'' of the first and fourth blocks of
  Figure~\ref{fig:ArticlePowerLawExponents}.
\item ``Volume outliers'' -- the names \textbf{which appear the most
  frequently} in the news, by being either in the top 1000 most
  frequent names in some year, or, separately, names in the top 0.1\%,
  as per Section~\ref{sec:FilteringNames}.  The graphs for these
  subsets of names are shown in Figures
  \ref{fig:DsArticlesConstTeSpikeyPeaks}
  and~\ref{fig:DsArticlesFracTeSpikeyPeaks} for the spike method, and
  Figures \ref{fig:DsArticlesConstTeLSPeaks}
  and~\ref{fig:DsArticlesFracTeLSPeaks} for the continuity method, and
  the statistical measurements appear in blocks 2, 3, 5 and 6 of
  Figure~\ref{fig:ArticlePowerLawExponents}.
\end{itemize}

From the 1900's to the 1940's, the fame durations in both categories
of outliers do tend to decrease, with the decreases across that time
interval statistically signicantly lower-bounded by 1-2 weeks via
99\% bootstrapping intervals. Heuristically, this seems consistent
with our original hypothesis that accelerating communications shorten
fame durations: 1-2 weeks is a reasonable delay to be incurred by
sheer communications delay before the omnipresence of telegraphy and
telephony. We note with curiosity that this effect applies only to the
highly-famous outliers rather than the typical fame durations. We
posit that this is perhaps due to median fame durations being
typically attributable to people with only geographically localized
fame, which does not get affected by long communication delays. We
leave to further work a more nuanced study to test these hypotheses
around locality and communication delays affecting news spread in the
early 20th century.

After the 1940's, on the other hand, we see no such decrease. On the
contrary, the durations of fame for both the duration outliers and the
volume outliers reverse the trend, and actually begin to slowly
increase.  Using the bootstrapping method, per
Section~\ref{sec:StatMeas}, we get the results marked in red in
Figure~\ref{fig:ArticlePowerLawExponents}: in almost all of the
outlier studies\footnote{7 out of the 8 outlier studies show
  statistically significant increases between the 1940's and the
  1980's, and between the 1960's and the 2000's. The sole exception is
  the 90th percentile of the spike method. Given that the bootstrap
  values in that experiment, discretized to whole weeks, range between
  3 and 4 weeks, we don't consider it surprising that the increases
  there were not measured to be significant by 99\% bootstrap
  intervals.}, we see that the increase in durations is statistically
significant over 40-year gaps for both categories of fame
outliers. For example, the median fame duration according to
continuity peaks for the top 1000 names (50th percentile column of the
fifth block) appears as ``27 (25 .. 29)'' in the period 1945-9 and ``52 (49
.. 56)'' for the period 1985-9: with 99\% confidence, the median
duration was less than 29 days in the former period, but greater than
49 days in the latter.

We also ran experiments for names that have outlier durations
\textit{within} the subset of names with outlier volumes. The same
general trends were seen there as with the above outlier studies, but,
with a far shallower pool of data, the bootstrapping-based error bars
were generally large enough to not paint a convincing, statistically
significant picture.

\paragraph{Power law fits}

The column titled ``power law exponent'' in
Figure~\ref{fig:ArticlePowerLawExponents} shows the maximum likelihood
estimates of the power law exponents for various five-year-long peak
periods.  We focus on the first and fourth blocks, which show the
estimates for the full set of names for the spike method and the
continuity method respectively.

For both peak methods, the fitted power law exponents remain in fairly small
ranges --- between -2.77 and -2.45 for continuity peaks, and between
-2.63 and -2.32 for spike peaks. In Figures
\ref{fig:AllArticlesLSPeaks} and \ref{fig:AllArticlesSpikeyPeaks} we
show the actual distributions, and, for reference, comparisons with
the power-law fit for the 2005-2009 data (a straight line on these
log-log plots).

Furthermore, the continuity peaks fits also support the above
observation of slowly-growing long-tail fame durations from 1940
onward. That is, power-law exponents from 1940 onward slowly move
toward zero, with statistically significant changes when compared at
40-year intervals. The fluctuations and the error bars for both
methods are rather noticeable, though, suggesting that power laws make for
only a mediocre fit to this data.

\section{Results: Blog Posts}

\label{sec:BloggerResults}

\begin{figure*}[t]
\noindent
\includegraphics[width=0.33\textwidth]{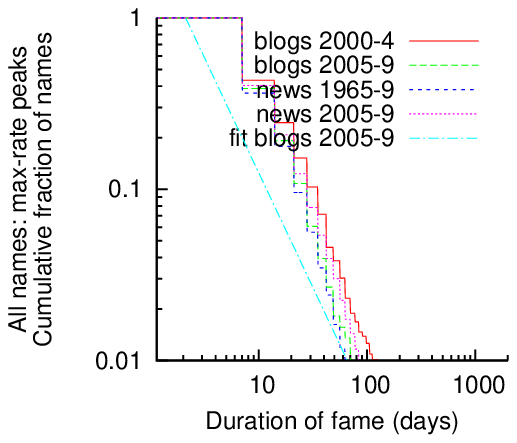}
\includegraphics[width=0.33\textwidth]{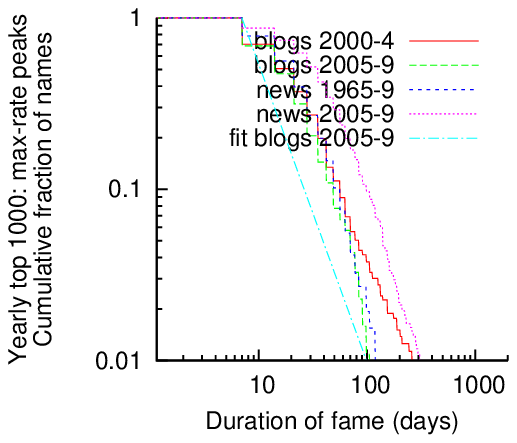}
\includegraphics[width=0.33\textwidth]{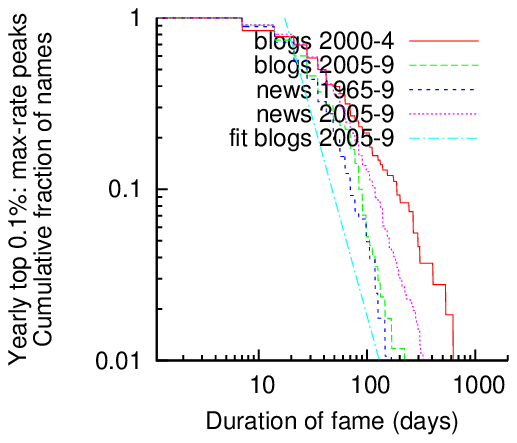}\\
\noindent
\includegraphics[width=0.33\textwidth]{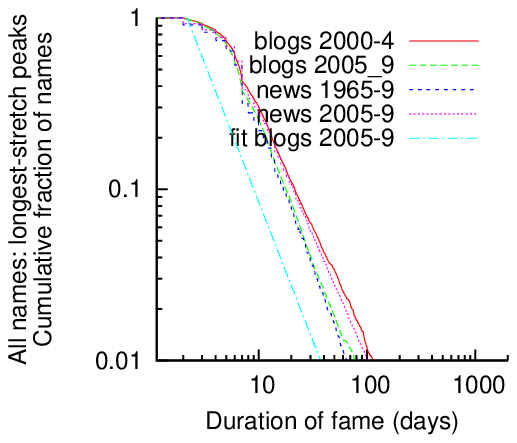}
\includegraphics[width=0.33\textwidth]{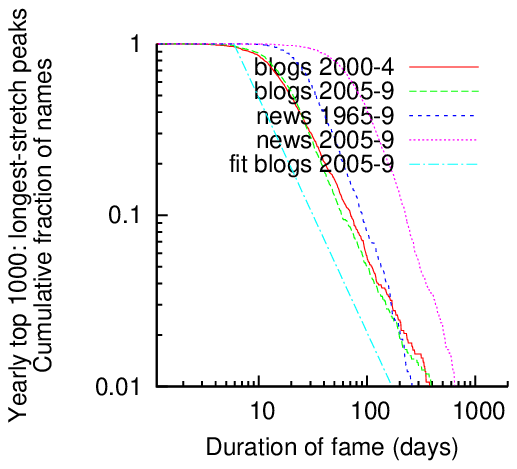}
\includegraphics[width=0.33\textwidth]{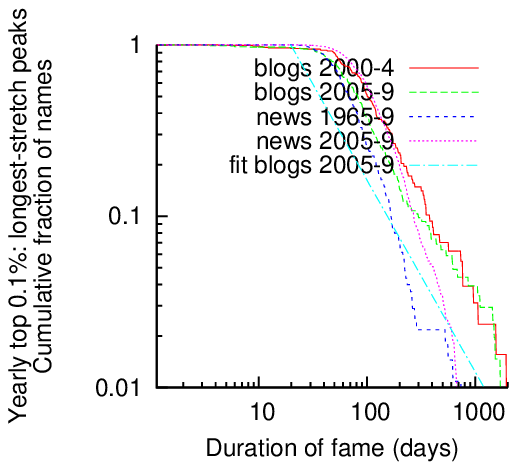}
\caption{
  \label{fig:BlogCumulativeGraphs}
  Cumulative duration-of-fame graphs for the blog corpus.
  The graphs at the top show the spike method results (for all
  names, top 1000, and top 0.1\%), and those at the bottom show the
  continuity method results.}
\end{figure*}

We also ran our experiments on a second set of data consisting of public
English-language blog posts from the Blogger service.  We began by
sampling so that the number of blog posts in each month in our data
set was equal to the number of news articles we sampled in each month,
as per Sec.~\ref{sec:Subsampling}.  The cumulative
graphs of fame duration from six experiments are shown in
Fig.~\ref{fig:BlogCumulativeGraphs}.  We combine the two methods for
identifying periods of fame with three sets of names described in
Section~\ref{sec:FilteringNames}.  The respective distributions from the
news corpus are superimposed for comparison.

The graphs of fame duration measured using the continuity method are
much smoother for the blog corpus than for the news corpus.  This
happens because whereas we only know which day each news article was
written, we know the time of day each blog entry was posted.

The continuity-method graphs (bottom of Figure~\ref{fig:BlogCumulativeGraphs})
had a distinctive rounded cap which surprised us at first.  We believe
it is caused by the following effect.  Peaks with only two mentions in
them are fairly common, and have a simple distinctive distribution
that is the difference between two sample dates conditioned on being
less than a week apart.  Since two dates that are longer than one week
apart cannot constitute a longest-stretch peak, the portion of the
graph with durations longer than one week does not include any names
from this two-sample distribution, and so it looks different.  Our
estimates of power-law exponents only consider the longest 20\% of
durations, so they ignore this part of the graph.

The estimates we computed for the power-law exponents of the duration
distributions for blog data are shown in
Figure~\ref{fig:BloggerPowerLawExponents}, and can be compared to the
figures for news articles in
Figure~\ref{fig:ArticlePowerLawExponents}.

The medians for both blogs and news for both methods are remarkably
the same, with no statistically significant differences. The power law
fits are also quite similar, although they show enough variation to
produce statistically significant differences. Qualitatively, we take
these as evidence that the fame distributions in news and blogs are
coarsely similar, and that it is not unreasonable to consider these
results as casting some light on more fundamental aspects of human
attention to and interest in celebrities, rather than just on the
quirks of the news business.

We do leave open the question of accounting for the occasionally
significant distinctions between outlier results for blogs, as
compared to news, especially for outlier-volume continuity peaks.

\section{Related Work}

Michel \emph{et al.}~\cite{ngrams} study a massive corpus of digitized
content in an attempt to study cultural trends.  The corpus they study
is even larger than ours in terms of both volume and temporal
extension.

Leetaru~\cite{leetaru2011culturomics} presents evidence that sentiment analysis
of news articles from the past decade could have been used to predict the
revolutions in Tunisia, Egypt and Libya.

Our spike method for identifying periods of fame is motivated in part
by the work of Yang and
Lescovec~\cite{yang2011patterns} on identifying patterns of temporal
variation on the web.  Szabo and Huberman~\cite{szabo} also consider
temporal patterns, in their case regarding consumption of particular
content items.  Kleinberg studies other approaches to
identification of bursts~\cite{kleinberg-bursts}.  

Numerous works have studied the propagation of topics through online
media.  Leskovec \emph{et al.}~\cite{memes} develop techniques for tracking
short ``memes'' as they propagate through online media, as a means to
understanding the news cycle.  Adar and Adamic~\cite{eytan-blogs}, and
Gruhl \emph{et al.}~\cite{idib} consider propagation of information
across blogs.

Finally, a range of tools and systems provide access to personalized
news information; see Gabrilovich et al~\cite{newsjunkie} and the
references therein for pointers.

\section{Acknowledgements}

The authors would like to thank Zoran Dimitrijevic and the Google News Archive
team for their help with the data; Danny Wyatt, Ed Chi, and Rachel
Schutt for statistical advice; and the anonymous reviewers for helpful
suggestions.

\bibliographystyle{plain}
\bibliography{bib}{}

\end{document}